\documentclass[aps,twocolumn,numerical]{revtex4}
\usepackage{color,graphicx,epsfig}
\usepackage{ifpdf}
\usepackage{amsmath}
\usepackage{bm}
\usepackage{color}
\usepackage[english]{babel}
\usepackage{graphicx}
\usepackage{amsfonts}
\usepackage{amssymb}
\usepackage{braket}
\usepackage{slashed}
\usepackage{enumerate}
\usepackage{mlmodern}
\definecolor{indigo}{rgb}{0.0, 0.25, 0.42}
\usepackage[colorlinks,allcolors=indigo]{hyperref}

\newcommand{\e}[1] {\epsilon^{#1}} 
\newcommand{\mnc}{m_{\slashed{C}}}
\newcommand{\cll}[2]{\mathcal{C}^{#1}_{LL#2}}
\newcommand{\sij}[2]{s_{{#1} {#2}}}

\newcommand{\nn}{\nonumber \\}
\newcommand{\mubar}{\bar{\mu}}

\newcommand{\amp}[2]{\mathcal{A}^{#1}_{#2}}

\newcommand{\angspinor}[1]{|#1\rangle}
\newcommand{\sqspinor}[1]{|#1]}

\newcommand{\AsqNLP}[2]{	\left[\mathcal{A}^{#1}_{#2}\right]^2_{\text{NLP}}}
\newcommand{\slog}{\log\left( \frac{s_{45}}{\mubar^2}\right )}

\newcommand{\npol}{\slashed{\Sigma}}
\newcommand{\pol}{\Sigma}

\newcommand{\dsigma}[2]{\left.	s_{12}^2\frac{d^2\sigma^{#1}_{#2}}{ds_{13}\,ds_{23}}\right|_{\text{NLP-LL}}}

\usepackage{xcolor}

\begin{document}

\title{Universality at next-to-leading power for jet associated processes}
\author{Sourav Pal}
\email{sourav@niser.ac.in}
 \affiliation{School of Physical Sciences, National Institute of Science Education and Research, \\
 An OCC of Homi Bhaba National Institute, Jatni 752050, India }
\author{Satyajit Seth}%
 \email{seth@prl.res.in}
\affiliation{Theoretical Physics Division, Physical Research Laboratory, \\ Navrangpura, Ahmedabad 380009, India} 
\date{\today}

\begin{abstract} The study of cross-sections in the threshold limit at next-to-leading power has been a subject of sustained interest for many years. We demonstrate the universality of leading logarithms at next-to-leading power for the production of arbitrary massive colourless particles in association with a jet, contingent upon the identification of appropriate combination of helicity configurations.
\end{abstract}

\maketitle


\section{Introduction}
\label{sec:intro}
The avalanche of data generated by the Large Hadron Collider (LHC), combined with the absence of compelling new physics signatures, necessitates a precise understanding of the well-established Standard Model. There are two primary approaches to enhance the accuracy of theoretical predictions: (i) calculating cross-sections at higher orders in perturbation theory, and (ii) performing resummation of certain large logarithmic terms to all orders in perturbation theory. For all physical processes occurring at the LHC, the differential cross-section in the threshold limit takes the following form:
\begin{align}
	\frac{d\sigma}{d\xi}\,\approx\,\sum_{n=0}^{\infty}\alpha_{s}^{n}& \bigg\{
		\sum_{m=0}^{2n-1}D^{\text{LP}}_{nm}\left(\frac{\log^{m}\xi}{\xi}\right)_{+}
		+d_{n}\delta(\xi)\nn
		&+\sum_{m=0}^{2n-1}D^{\text{NLP}}_{nm}\,\log^{m}\xi\bigg\}
	\,,\label{eq:gen-threshold}
\end{align} 
where $\xi$ is the threshold variable. In this equation, the first set
of logarithms and the delta function arise when one considers leading power
approximation of the soft radiation. These set of logarithms are universal due
to the factorisation properties of soft and collinear radiations. The
groundbreaking works of
refs.~\cite{Parisi:1980xd,Curci:1979am,Sterman:1987aj,Catani:1989ne,Catani:1990rp,Gatheral:1983cz,Frenkel:1984pz,Sterman:1981jc},
based on diagrammatic methods, were instrumental in developing methods for leading power (LP) resummation. 
Subsequently, several alternative methods of LP resummation were developed, including those based on Wilson lines~\cite{Korchemsky:1993xv,Korchemsky:1993uz},
renormalisation group (RG)~\cite{Forte:2002ni} and Soft Collinear Effective
Theory (SCET)~\cite{Becher:2006nr,Schwartz:2007ib,Bauer:2008dt,Chiu:2009mg}. A
comparison of these various resummation techniques of LP logarithms can be
found in ~refs.~\cite{Luisoni:2015xha,Becher:2014oda,Campbell:2017hsr}). The
second set of logarithms in eq.~\eqref{eq:gen-threshold} originates due to next-to-leading
power (NLP) approximation of the radiated particle's momentum, which is the focus of this article.  

The structure of scattering amplitudes was first studied almost sixty
years  ago~\cite{Low:1958sn,Burnett:1967km,DelDuca:1990gz} and continues to be an area of active research.  
Given the significant numerical impacts of the NLP logarithms~\cite{Kramer:1996iq,Ball:2013bra,Bonvini:2014qga,Anastasiou:2015ema,Anastasiou:2016cez,vanBeekveld:2019cks,vanBeekveld:2021hhv,Ajjath:2021lvg}, substantial progress has been made towards understanding these logarithms through the infrared structure of gauge theory amplitudes. 
Over the past decade, several methods have been developed to study these
logarithms~\cite{Grunberg:2009yi,Soar:2009yh,Moch:2009hr,Moch:2009mu,Laenen:2010uz,Laenen:2008gt,deFlorian:2014vta,Presti:2014lqa,Bonocore:2015esa,Bonocore:2016awd,Bonocore:2020xuj,Gervais:2017yxv,Gervais:2017zky,Gervais:2017zdb,Laenen:2020nrt,DelDuca:2017twk,vanBeekveld:2019prq,Bonocore:2014wua,Bahjat-Abbas:2018hpv,Ebert:2018lzn,Boughezal:2018mvf,Boughezal:2019ggi,Bahjat-Abbas:2019fqa,Ajjath:2020ulr,Ajjath:2020sjk,Ajjath:2020lwb,Ahmed:2020caw,Ahmed:2020nci,Ajjath:2021lvg,Kolodrubetz:2016uim,Moult:2016fqy,Feige:2017zci,Beneke:2017ztn,Beneke:2018rbh,Bhattacharya:2018vph,Beneke:2019kgv,Bodwin:2021epw,Moult:2019mog,Beneke:2019oqx,Liu:2019oav,Liu:2020tzd,Boughezal:2016zws,Moult:2017rpl,Chang:2017atu,Moult:2018jjd,Beneke:2018gvs,Ebert:2018gsn,Beneke:2019mua,Moult:2019uhz,Liu:2020ydl,Liu:2020eqe,Wang:2019mym,Beneke:2020ibj,vanBeekveld:2021mxn,Das:2024pac,Das:2025wbj,Bhattacharya:2025rqk,vanBijleveld:2025ekz,Czakon:2023tld,Agarwal:2025dvo,Hou:2025ovb}.
However, none of the existing methods can conclusively confirm the factorisation of
next-to-soft radiation at all orders for all collider processes in perturbation theory. 
While the universality of NLP logarithms has been established in case of colour singlet particle production~\cite{DelDuca:2017twk}, no unique resummation formula exists for coloured particles in the final state.  One approach to
recognise the patterns of NLP logarithms and develop a general resummation
formula involves analysing multiple processes with coloured particles in the
final state. In this context, the NLP logarithms for
prompt photon plus jet production were investigated
in~\cite{vanBeekveld:2019prq}. Additionally, attempts have been made towards
calculating these logarithms for vector boson plus jet
production~\cite{Sterman:2022lki,Boughezal:2019ggi,vanBeekveld:2023gio}.
However, due to the complex nature of the phase space integration, none of these
references provide complete expressions of the NLP logarithms.   

To remove the complexity and streamline the NLP
calculations, we developed a straightforward formalism in
ref.~\cite{Pal:2023vec} to obtain NLP
corrections due to the radiation of gluons.
Our formalism combines colour-ordered helicity amplitudes and
soft theorems of gauge theory~\cite{Strominger:2013jfa,Casali:2014xpa,Luo:2014wea}. 
Since soft theorems apply only to the emission of gluons, we could
use them only to compute the NLP logarithms that arise due to the next-to-soft gluon
radiation. Further, to account for soft quark contributions at NLP, we defined
soft quark operators in terms of helicity spinors in ref.~\cite{Pal:2024eyr}.
Using these two approaches, we derived the NLP threshold
corrections for all possible partonic channels in Higgs plus one jet
production. Moreover,  based on symmetry considerations and the similarity in structure between the colour-ordered
amplitudes of Higgs plus one jet and pseudo-scalar Higgs plus one jet
production, we established that these logarithms are universal for both processes. 

The next natural step in investigating the universal structure of scattering cross-sections at NLP accuracy is to calculate the NLP corrections for the production of a massive vector boson associated with a jet at the LHC. Prior to undertaking explicit computations of these processes, we propose in this article that the coefficients of NLP leading logarithms exhibit a universal analytic structure, provided the appropriate combination of helicity configurations is identified, for the production of any massive color-singlet particle -- regardless of its spin -- in association with a jet. These coefficients can be derived through the mass factorisation formula, using the helicity-dependent Altareli-Parisi splitting functions for both the quark and the gluon.

\section{Next-to-soft gluon and Soft quark}
It is well established that soft gluon radiation is the sole source of threshold logarithms at leading power. 
In contrast, next-to-leading power (NLP) logarithms receive contributions from
two distinct sources: $(i)$ next-to-soft gluon radiation and $(ii)$ soft quark radiation. 
In this section, we describe the methods introduced 
in~\cite{Pal:2023vec,Pal:2024eyr} for computing these contributions.
To capture the next-to-soft gluon radiation effects, we use the soft theorem of gauge theory~\cite{Strominger:2013jfa,Luo:2014wea,Casali:2014xpa}, which relates $(n+1)$-particle helicity amplitude to the $n$-particle amplitude at LP and NLP as,    
\begin{align}
	& \mathcal{A}_{\,n+1}^{\text{LP+NLP}}\bigg(\left\{ \lambda\angspinor
	s,\sqspinor s\right\} ,\left\{ \angspinor 1,\sqspinor 1\right\}
,\ldots,\left\{ \angspinor n,\sqspinor n\right\} \bigg)  \,=\,
	\nn & \frac{1}{\lambda^{2}}\frac{\braket{1n}}{\braket{1s}\braket{ns}}\times\mathcal{A}_{\,n}\bigg(\left\{ \angspinor 1,\sqspinor{1^{\prime}}\right\} ,\ldots,\left\{ \angspinor n,\sqspinor{n^{\prime}}\right\} \bigg)\,, \label{eq:LPplusNLP}
\end{align}
where we consider the emission of a gluon with helicity  \lq$+$\rq\, and use the holomorphic soft limit, $|s\rangle \to \lambda |s\rangle$. The amplitude on the right-hand side of the above equation corresponds to the simple non-radiative $n$-particle amplitude, with a pair of square spinors being shifted as,  
\begin{equation}
	\sqspinor{1^{\prime}}\,=\,\sqspinor 1+\Delta_{s}^{(1,n)}\sqspinor s\,,\nn\sqspinor{n^{\prime}}\,=\,\sqspinor n+\Delta_{s}^{(n,1)}\sqspinor s\,,
	\Delta_{s}^{(i,j)}=\lambda\frac{\braket{js}}{\braket{ji}}\,.
	\label{eq:gen-shifts}	
\end{equation}
In the case where the radiated gluon carries negative helicity, the square spinors are naturally replaced by their corresponding angle spinors.

In order to compute the soft quark contribution using colour-ordered helicity amplitudes, we define two soft quark operators in~\cite{Pal:2024eyr}. These operators are constructed by combining a soft quark with its neighbouring hard colour particle. This combination yields an effective hard particle which subsequently participate in the corresponding non-radiative amplitudes. As a result, the $ (n+1)$-particle colour-ordered amplitude under soft quark emission is given by~\cite{Pal:2024eyr},
\begin{align}
    \amp{s^{i} h^{j}}{n+1}\,=\,\mathcal{Q}^{s^{i} h^{j}\to c^{k}}\,\amp{c^{k}}{n}\,,\qquad
    \label{eq:sqfeynman}
\end{align}
where $\mathcal{Q}$ is the soft quark operator, $s^i$ denotes the ``{\em soft}\,'' (anti-)quark with helicity $i$; $h^j$ is the ``{\em hard}\,'' particle with helicity $j$ and the ``{\em clubbed}\,'' colour particle $c^k$ with helicity $k$ is being formed by combining the hard and soft particles of the $(n+1)$-particle amplitude. 
Since the soft quark effect is fully captured within the operator, the helicity and momentum of the resulting clubbed particle are taken to be the same as those of the original hard particle {\em i.e.,} $j=k$ in the equation above. There are two distinct ways to merge a soft quark with a hard particle, resulting into the following two fundamental non-zero soft quark operators~\cite{Pal:2024eyr},
\begin{align}
	\mathcal{Q}^{q_{s}^{+}\bar{q}_{h}^{-}\to g_{c}^{-}}\,=  \,  -\frac{1}{\braket{sh}}\,, \qquad  
\mathcal{Q}^{g_{h}^{+} q_{s}^{+}\to q_{c}^{+}}\,=  \, -\frac{1}{\braket{sh}}\,, 
\label{eq:sqop}
\end{align}
and all other possible combinations can be derived from these two.

\section{NLP Corrections}
In this article, we focus on the processes that involve production of one
colourless particle associated with a jet at the LHC.  
The colourless particle can be either massive or massless, and 
the general form for these processes can be written as, 
\begin{align}
	P^a_C (p_1)+ P^b_C (p_2) \rightarrow P_{\slashed{C}} (p_3) + P^d_C (p_4) \, , 
\label{eq:gen2to2}	
\end{align}
where the suffix $C$ and $\slashed{C}$ denote coloured and colourless
particles ($P$), respectively. 
In a completely general setup $p_3^2=\mnc^2$, where 
in the Standard Model $\mnc\,\in\,\{0_{\gamma}, m_W, m_Z, m_H\}$. 
Our interest lies in studying the threshold logarithms arising from the emission of an additional coloured particle with momentum $p_5$, leading to the following scattering process,  
\begin{align}
	P^a_C (p_1)+ P^b_C (p_2) \rightarrow P_{\slashed{C}} (p_3) + P^d_C (p_4)  + P^e_C (p_5)\, .  
\label{eq:gen2to3}	
\end{align}
Here $p_5$ can either be a quark, an anti-quark or a gluon. 
Since we are interested in obtaining scattering cross-section up to NLP, we may express the scattering amplitude as, 
\begin{align}
	\mathcal{A}\,=\, \mathcal{A}_\text{LP}+\mathcal{A}_\text{NLP} \,,
\end{align}
and squaring this amplitude we obtain,
\begin{align}
	\mathcal{A}^2\,=\,\mathcal{A}_\text{LP}^2+2\text{Re} \left (\mathcal{A}_\text{NLP}\mathcal{A}_{\text{LP}}^\dagger \right ) \,,
\end{align}
where the term $\mathcal{A}_\text{NLP}^2$ is being neglected, as it starts contributing at the next-to-next-to leading power.
Throughout this article, we use the shorthand notation $\AsqNLP{}{}$ to represent the second term. 
In terms of holomorphic scaling of spinors, the colour-ordered
helicity amplitudes and their squared parts scale at NLP as, 
\begin{align}
	\mathcal{A}_{\text{NLP}}\,\approx \, \frac{1}{|5\rangle}\,, \;
	\AsqNLP{}{} \approx \frac{1}{\sij{i}{5}} \,,
\label{eq:genlpnlp}	
\end{align}
where $\sij{i}{5} = \{\sij{1}{5},\sij{2}{5},\sij{4}{5}\}$, since the momentum of the massive particle can always be expressed using four-momentum conservation as  $p_3=p_1+p_2-p_4-p_5$. 
Once the squared amplitudes for each independent helicity configuration are obtained, the next step is to perform the phase space integration over the unobserved parton's momentum. To deal with the infrared singularities, we choose the space time dimension $d=(4-2\epsilon)$, and parameterize the particle momenta in the rest frame of $p_4$ and $p_5$ as, 
\begin{eqnarray}
	p_1&=&(E_1,0,\cdots,0,E_1) \,, \nonumber \\
	p_2&=& (E_2,0,\cdots,0,p_3\sin \psi, p_3\cos \psi - E_1) \,, \nonumber\\
	p_3&=&-(E_3,0,\cdots,0,p_3\sin \psi, p_3\cos \psi ) \,, \nonumber \\
	p_4&=&-\frac{\sqrt{s_{45}}}{2} (1,0,\cdots,0,\sin \theta_1 \sin \theta_2,\sin\theta_1\cos \theta_2,\cos \theta_1) \,, \nonumber\\
	p_5&=&-\frac{\sqrt{s_{45}}}{2} (1,0,\cdots,0,-\sin \theta_1\sin \theta_2,-\sin\theta_1\cos \theta_2,
	\nn & & -\cos \theta_1)\,. 
	\label{eq:para1} 
\end{eqnarray}
With this phase space parametrisation, the helicity dependent differential cross-section at NLP is given by~\cite{Pal:2023vec,Pal:2024eyr}, 
\begin{align}
	\left.	s_{12}^2\frac{d^2\sigma^{h_1 h_2 h_3 h_4 h_5}}{ds_{13}ds_{23}}\right|_{\text{NLP}}
	\,&= \, \mathcal{F}_{ab} \left (\frac{s_{45}}{\mubar^2}\right )^{-\epsilon}\,\overline{\mathcal{A}_{\text{NLP}}^2} \,,
	\label{eq:crossx}	
\end{align}
where
\begin{align}
	\mathcal{F}_{ab}\,&=\, \frac{1}{2}K_{ab}\,G^2\left( \frac{\alpha_{s}(\mubar^2)}{4\pi}\right )^2\,
	\left(\frac{s_{13}\,s_{23}-m_H^2\,s_{45}}{\mubar^2\,s_{12}}\right
		)^{-\epsilon} \,, \nn 
\end{align}
and
\begin{align}
	\overline{\mathcal{A}_\text{NLP}^2}\,=&\,\int_{0}^{\pi} d\theta_1\, (\sin\theta_1)^{1-2\epsilon}\int_{0}^{\pi}d\theta_2\, (\sin\theta_2)^{-2\epsilon} \AsqNLP{}{} \,.
\end{align}
Note that $K_{ab}$ is a factor related to the colour and spin average of
the initial state particles, and $G$ is the coupling constant associated with the massive particle. Since we are
only interested in the leading logarithms at NLP, therefore we can treat 
$\mathcal{F}_{ab}$ as independent of the angular integrations. However, for next-to-leading logarithms at NLP, one would need to expand this factor around the threshold variable, though this is not necessary for the current purpose. 
The general form of the scattering cross-section expanded around $\epsilon$ after performing the phase
space integration, is given by, 
\begin{align}
	& \dsigma{h_1 h_2 h_3 h_4 h_5}{} \, \, \nn= & \left\{\mathcal{C}_{-1}\frac{1}{\e{}}  +
	\mathcal{C}_{45}\frac{1}{\left(\sij{4}{5}\right)_+} +\cll{}{} \slog
\right\}
(\amp{h_1h_2h_3h_4}{})^2 \,,
\label{eq:gensigmaNLP}	
\end{align}
where $\mathcal{C}$s' represent the coefficients of the corresponding factors that are multiplied by them.

\section{Universality}
Infrared singularities in the scattering cross-section appear as divergences that manifest as poles in $\epsilon$ following dimensional regularisation with $d=(4-2\epsilon)$. The singularity that appears in the differential cross-section at NLP, as shown in eq. \eqref{eq:gensigmaNLP}, is cancelled when the effects of mass factorisation are taken into account, rendering the cross-section finite. This finiteness ensures that the coefficient $\mathcal{C}_{-1}$ can be determined from the mass factorisation contribution and is given by, 
\begin{align}
	\mathcal{C}_{-1}\,=&\, -\int_{0}^{1} \frac{dx_1}{x_1} \,\Gamma_{a^\prime\to a e}(x_1,\epsilon) \,\hat{s}_{12}^2\,\frac{d^2\sigma^{(0)}(\hat{s}_{12},\hat{s}_{13},s_{23})}{d\hat{s}_{13}ds_{23}} \frac{1}{(\amp{}{})^2} \nn 
& -\int_{0}^{1} \frac{dx_2}{x_2} \,\Gamma_{b^\prime\to b e}(x_2,\epsilon) \,\hat{s}_{12}^2\,\frac{d^2\sigma^{(0)}(\hat{s}_{12},s_{13},\hat{s}_{23})}{ds_{13}d\hat{s}_{23}} \frac{1}{(\amp{}{})^2} \,,
	\label{eq:mfreq}	
\end{align}	
where 
\begin{align}
        &(\amp{}{})^2\,=\,(\amp{h_1h_2h_3h_4}{})^2, \;
	\hat{s}_{12}\,=\,x_1\,x_2\, \sij{1}{2}\,,  \nn
	&\hat{s}_{13}\,=\,x_1(\sij{1}{3}-\mnc^2) +\mnc^2 \,, \nn 
	&\hat{s}_{23}\,=\,x_2(\sij{2}{3}-\mnc^2) +\mnc^2\,.
\end{align} 
$\sigma_0$ is the Born scattering cross-section and $\Gamma$ represents the helicity dependent splitting kernel defined as, 
\begin{align}
	\Gamma_{i\to jk}\,=\,-\frac{\alpha_{s}}{2\pi}\,\frac{1}{\epsilon}
	\mathcal{P}_{i\to jk}\,,
\end{align}
with $\mathcal{P}$ being the helicity dependent Altarelli-Parisi splitting functions. 
Note that $\mathcal{C}_{-1}$ can be determined, as outlined above, for the radiation of both next-to-soft gluons and soft quarks, irrespective of whether the production channel is diagonal ({\em i.e.,} $q\bar{q}$ or $gg$ initiated) or off-diagonal ({\em i.e.,} $qg$ or $\bar{q}g$ initiated).

Since the Eikonal factor appears as an overall multiplicative term in the expression for $\AsqNLP{}{}$, 
its contribution at NLP involves multiplication with a linear combination of $\sij{i}{5}$ with $i \in \{1, 2, 4\}$. This structure leads to six distinct types of terms, such as -- $\frac{1}{\sij{i}{5}}$ and $\frac{\sij{i}{5}}{\sij{\{i+1\}}{5} \sij{\{i+2\}}{5}} $, each accompanied by appropriate coefficients that are independent of $\sij{i}{5}$. Upon performing the angular integration of these terms and multiplying by the appropriate phase space factor, it follows straightforwardly that, 
\begin{align}
	\cll{}{}\,=\,-\,\mathcal{C}_{-1}\,.
\label{eq:clltoc1}	
\end{align}
This represents one of the central findings of this article, exhibiting that the coefficients of the NLP leading logarithms can be derived from the computation of mass-factorisation related contributions. Remarkably, this relation holds across both diagonal and off-diagonal production channels. Although next-to-soft gluon and soft quark radiation differ in their physical origin and necessitate distinct treatments for extracting NLP logarithms, they nonetheless demonstrate the same universal structure in the threshold logarithms. This striking consistency strongly points towards the presence of an underlying universality governing these logarithmic contributions.

The NLP leading logarithms exhibit a universal analytical structure for both next-to-soft gluon emission and soft quark radiation, provided that the relevant helicity configurations are properly accounted for. This universality persists irrespective of whether the final-state massive colourless particle is of spin-0 or spin-1. In the case of soft (anti-)quark emission, the NLP leading logarithms maintain a consistent analytical form when all contributing Born-level squared amplitudes are accurately identified -- specifically, those constructed by {\em clubbing} the soft (anti-)quark with the adjacent coloured particle in colour-ordered helicity amplitudes, along with a sum over all polarisations of the massive particle. Likewise, for next-to-soft gluon emission, the NLP leading logarithms exhibit universal analytical structures upon summation over the helicities of the emitted gluon and the polarisations of the massive particle.

In what follows, we consider two specific helicity configurations in the context of Higgs+jet production in the large top-mass limit -- one involving next-to-soft gluon emission and the other involving soft quark radiation: 
\begin{align}
	\cll{++\npol-+}{,\,1_g\,2_q\,3_H\,4_{\bar{q}}\,5_{g(s)}}&=\,
	4 \pi \mathcal{F}_{qg}\, \bigg\{C_A \left(\frac{4}{s_{23}}\right)
	\nn & +2 C_F \left(\frac{1}{s_{23}}\right) \bigg\}
	\left|\amp{++\npol-}{1_g\,2_q\,3_H\,4_{\bar{q}}}\right|^2 \,,
	\label{eq:Hsg}	
\end{align}
\begin{align}
	\cll{-+\npol-+}{,\,1_{g}\,2_q\,3_H\,4_{\bar{q}(s)}\,5_{g}}&=\,
	2 \pi \mathcal{F}_{qg}\, \bigg\{C_A \left(\frac{2}{s_{23}}\right)  
 	\nn & + 2 C_F \left(\frac{1}{\sij{2}{3}}\right)  \bigg\} \left|\amp{-+\npol+}{1_{\bar{q}}2_q\,3_H\,5_g}\right|^2  \,.
\label{eq:Hsq}
\end{align}
In each case, the soft particle is denoted by a ``$(s)$" in the subscript. The quantities $C_A=N$ and $C_F=\frac{N^2-1}{2N}$ represent the quadratic Casimir invariants associated with the adjoint and fundamental representations of the $SU(N)$ colour group, respectively. Here $\npol$ signifies that the Higgs boson, being a scalar particle, does not possess any polarisation degrees of freedom. Note that $\cll{++\npol--}{,\,1_g\,2_q\,3_H\,4_{\bar{q}}\,5_{g(s)}}=0$, as the maximally helicity violating amplitudes do not contribute to this process at NLP.  

Based on the preceding discussion, it becomes evident that one can anticipate the structure of NLP leading logarithms for analogous processes, such as $W$+jet production. These are expected to follow the form as given below, 
\begin{align}
	&\left[\cll{++\pol-+}{,\,1_g\,2_{q}\,3_W\,4_{q^\prime}\,5_{g(s)}} +\; \cll{++\pol--}{,\,1_g\,2_{q}\,3_W\,4_{q^\prime}\,5_{g(s)}}\right] \nn
	&=
	4 \pi \mathcal{F}_{qg}\, \bigg\{C_A \left(\frac{4}{s_{23}}\right)
	 + 2 C_F \left(\frac{1}{s_{13}}\right) \bigg\}
	\left|\amp{++\pol-}{1_g\,2_{q}\,3_W\,4_{q^\prime}}\right|^2 \,,
	\label{eq:Wsg}	
\end{align}
\begin{align}
	\cll{-+\pol-+}{,\,1_{g}\,2_q\,3_W\,4_{q^\prime(s)}\,5_{g}}&=\,
	2 \pi \mathcal{F}_{qg}\, \bigg\{ 2 C_F \left(\frac{1}{\sij{2}{3}}\right)
\left|\amp{-+\pol+}{1_{q^\prime}2_q\,3_W\,5_g}\right|^2 \bigg\} \,.
\label{eq:Wsq}
\end{align}
The notation $\pol$ in the superscript indicates a summation over the polarization states of the $W$ boson is being carried out. Between $q$ and $q^\prime$, one denotes a quark and the other an anti-quark, with the assignment depending on the electric charge of the $W$ boson. It is noteworthy that no term proportional to $C_A$ appears in eq.~\eqref{eq:Wsq}, as there is no $gg$ initiated channel contributing to $W$+jet production at the Born level.  
\begin{figure}[t]
	\begin{center}
	\includegraphics[scale=0.33]{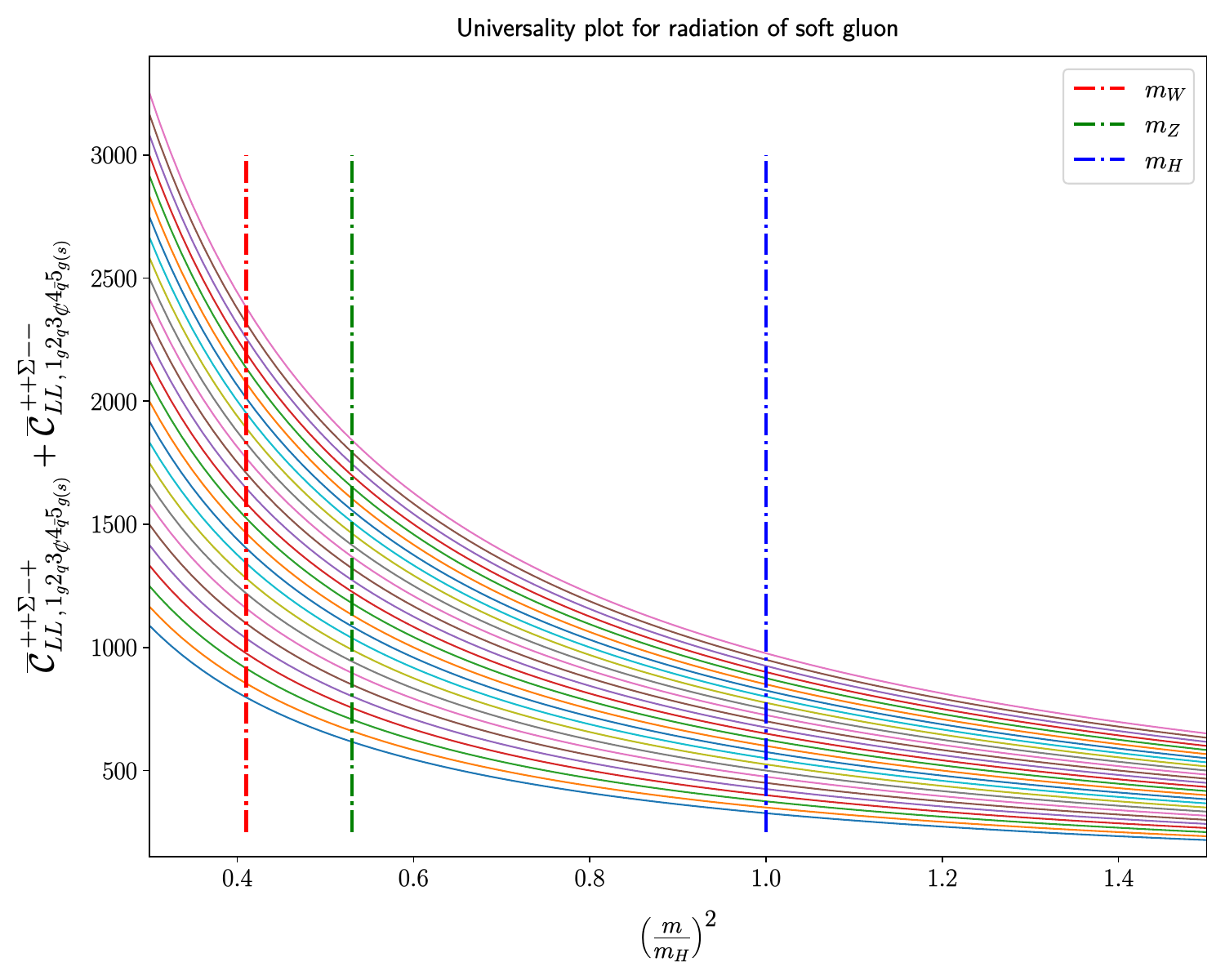}
	\end{center} 
	\caption{The plot illustrates the variation of the sum of $C_{LL}$s' as in eq.~\eqref{eq:Wsg}, with $\mathcal{F}$ and $|\mathcal{A}|^2$ excluded, for a generic colourless particle $\slashed{C}$, with its mass scaled to the Higgs mass and varied over a wide range across different phase space points. Discontinuous vertical lines in different colours indicate results for the $W$, $Z$ and Higgs bosons.}
\label{fig:plotsg}	
\end{figure}
\begin{figure}[tbh]
	\includegraphics[scale=0.33]{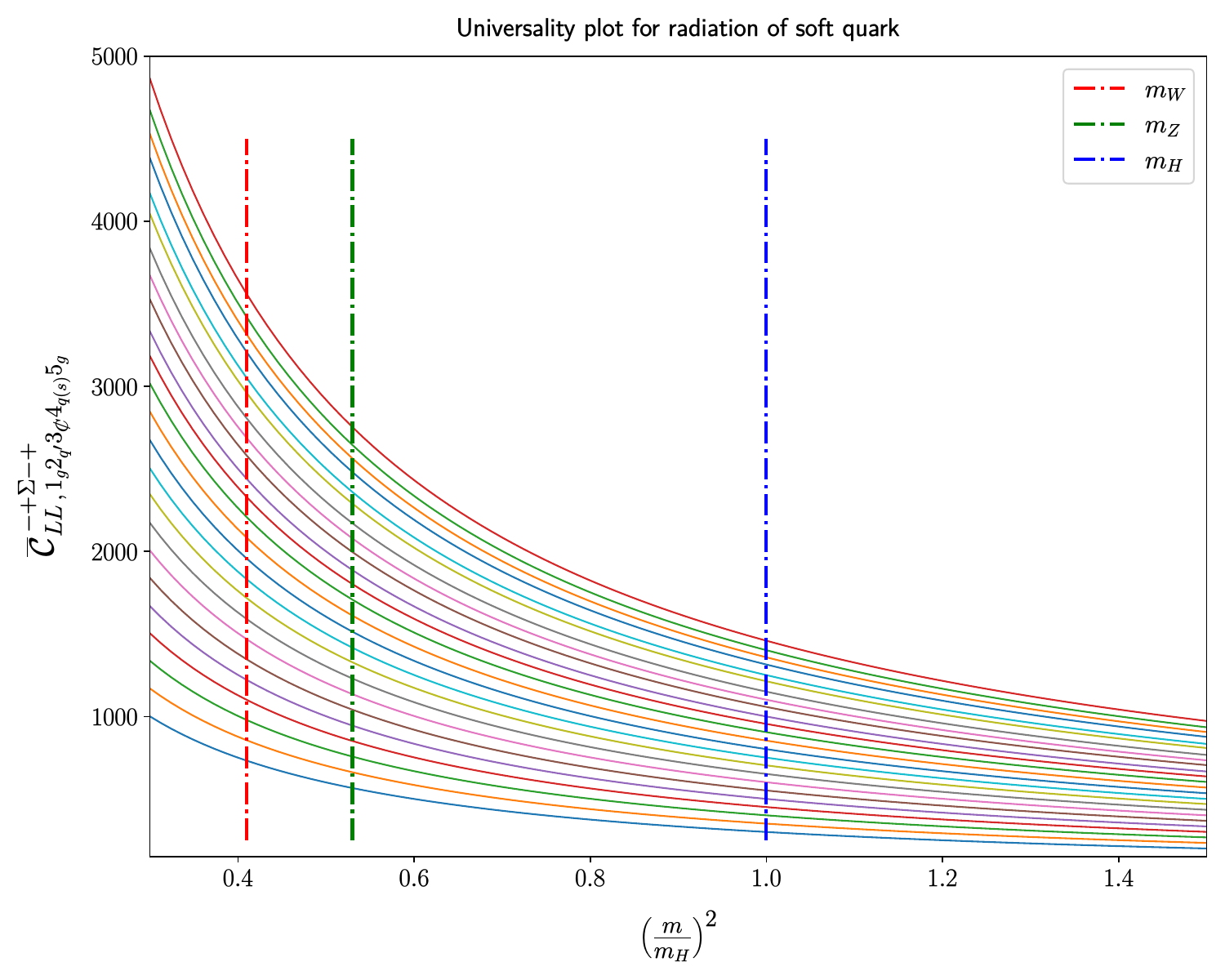} 
	\caption{The plot illustrates the variation of $C_{LL}$ as in eq.~\eqref{eq:Wsq}, with $\mathcal{F}$ and $|\mathcal{A}|^2$ factored out, for a generic colourless particle $\slashed{C}$, with its mass scaled to the Higgs mass and varied over a wide range across different phase space points. Discontinuous vertical lines in different colours represent the results for the $W$ and $Z$ bosons, as well as for the Higgs boson, with only the portion proportional to $C_F$.}
\label{fig:plotsq}	
\end{figure}
In Figures \ref{fig:plotsg} and \ref{fig:plotsq}, we present the NLP leading logarithmic coefficients, with the corresponding $\mathcal{F}_{ab}$ factors and squared amplitudes removed (thus, $C_{LL}$ is denoted as $\bar{C}_{LL}$ on the vertical axis), for helicity configurations analogous to those in eqs.~\eqref{eq:Wsg} and \eqref{eq:Wsq}, respectively, but applied to a generic colourless particle $\slashed{C}$. The mass of this particle is scaled to the Higgs boson mass $m_H$ and varied over a wide range for different phase space points. The universal nature of the analytic forms of these coefficients enables the identification of the NLP leading logarithmic values for the production of any colourless particles in association with a jet. We explicitly highlight the values for the $W$, $Z$ and Higgs bosons. By multiplying these values with the appropriate $F_{ab}$ factors and squared amplitudes, the actual NLP leading logarithmic values can easily be obtained. Note that from Figure \ref{fig:plotsq}, only the portion proportional to $C_F$ can be estimated for the Higgs+jet production, as eq.~\eqref{eq:Wsq} does not contain any $C_A$ term.
A complete and explicit calculation of the NLP contributions for $W^{\pm}$ and $Z$-boson production in association with a jet is beyond the scope of this article and will be presented in forthcoming studies, which will focus on identifying the NLP leading logarithms for individual helicity configurations and verifying the universality of the analytic structures discussed herein.

\section{Conclusion}
The increasing precision of experimental data continues to demand a deeper and more refined understanding of the Standard Model. In this context, the universality of NLP threshold logarithms has garnered considerable attention over the past decade, particularly in processes involving the production of a massive colourless particle in association with a jet. In this work, we provide a systematic and concise analysis of the universal structure of these logarithms, employing colour-ordered helicity amplitudes, that facilitate the identification of universal features.

We argue that the coefficients governing these universal logarithms can be determined through mass factorization using universally defined splitting kernels, as required by the finiteness of physical cross-sections. This universality holds regardless of the spin of the massive colourless particle, the nature of the associated jet (quark or gluon), or the production channel (diagonal or off-diagonal), provided the appropriate helicity configurations are identified. For soft quark emission, the NLP leading logarithmic contributions retain a consistent analytic structure, provided all relevant Born-level squared amplitudes are correctly included. These amplitudes correspond to configurations in which the soft quark is {\em clubbed} with the adjacent coloured particle in the colour-ordered helicity amplitudes, with a sum over the polarizations of the massive particle. Similarly, for next-to-soft gluon emission, the NLP leading logarithms display the universal analytic behaviour upon summation over both the gluon helicities and the polarizations of the massive particle. 

This article presents, for the first time, the most compelling demonstration of the universal features of NLP leading logarithms in the production of a colour-singlet particle in association with a jet. It offers a remarkably simple, practical and systematic framework for determining NLP leading-logarithmic coefficients for any massive colourless particle within the Standard Model and beyond. As such, this study sets a definitive foundation for the resummation of NLP leading logarithms in jet-associated production processes at the LHC.

\section*{Acknowledgements}
The research of SP is supported by the ANRF/SERB Grant SRG/2023/000591 and
NISER Plan Project No.: RIN-4001. The work of SS is supported by the Department of Space, Government of India. SS  acknowledges partial support by the ANRF/SERB MATRICS under Grant No. MTR/2022/000135.

\bibliographystyle{bibstyle}
\bibliography{Wjet}
\end{document}